\documentstyle[sprocl,epsfig]{article}

\newcommand{\feff   }{\mbox{$F_{\rm eff}^{\gamma}$}}
\newcommand{\qsq    }{\mbox{$Q^2$}}
\newcommand{\psq    }{\mbox{$P^2$}}
\newcommand{\wsq    }{\mbox{$W^2$}}
\newcommand{\qzm    }{\mbox{$\langle \qsq \rangle$}}
\newcommand{\pzm    }{\mbox{$\langle \psq \rangle$}}
\newcommand{\gev    }{\mbox{$\rm GeV$}}
\newcommand{\gevsq  }{\mbox{$\rm GeV^2$}}
\newcommand{\der    }{\mbox{${\mathrm d}$}}
\newcommand{\ft     }{\mbox{$F_{\mathrm{2}}^{\gamma}$}}
\newcommand{\ftxq   }{\mbox{$\ft(x,\qsq)$}}
\newcommand{\fl     }{\mbox{$F_{\rm L}^{\gamma}$}}
\newcommand{\flxq   }{\mbox{$\fl(x,\qsq)$}}
\newcommand{\aem    }{\mbox{$\alpha$}}
\newcommand{\aemsq  }{\mbox{$\alpha^2$}}
\newcommand{\invpb  }{\mbox{$\mathrm{pb}^{-1}$}}
\newcommand{\epem   }{\mbox{e$^+$e$^-$}}
\newcommand{\xvis   }{\mbox{$x_{\rm vis}$}}
\newcommand{\vvis   }{\mbox{$v_{\rm vis}$}}
\newcommand{\fpar   }{\mbox{$f^{\rm part}$}}
\def\be{\begin{equation}}
\def\ee{\end{equation}}
\def\bea{\begin{eqnarray}}
\def\eea{\end{eqnarray}}
\begin{document}
 \title{HADRONIC STRUCTURE FUNCTIONS OF THE PHOTON MEASURED AT LEP}
 \author{Richard Nisius}
 \address{\it{CERN, Email: Richard.Nisius@cern.ch}}
 \author{Daniel Haas}
 \address{\it{University of Basel, Email: Daniel.Haas@cern.ch}}
 \author{Claus Grupen}
 \address{\it{University of Siegen and CERN, Email: Claus.Grupen@cern.ch
        \footnote{Invited talks given at the DIS 2000
         Conference, 25-30 April 2000, Liverpool, 
         England, to appear in the Proceedings.}\\  
          }}
%
%
 \maketitle
 \abstracts{The measurements of hadronic structure functions of
 the photon based on the reaction 
 $\mathrm{e}\mathrm{e} \rightarrow \mathrm{e}\mathrm{e}
 \gamma^{(\star)}(\psq)\gamma^{\star}(\qsq)
 \rightarrow \mathrm{e}\mathrm{e}\,\mbox{\it hadrons}$ are discussed. 
 This review covers the latest developments in the analysis
 and the most recent measurements at LEP.}
%
%
\section{Introduction}
\label{sec:intro}
 One of the most powerful tools to investigate the structure of quasi-real
 photons, $\gamma$, is the measurement of photon structure functions
 in deep inelastic electron-photon scattering at electron-positron
 colliders. 
 A recent review on this topic  along with the references to the 
 published results can be found in~\cite{NIS-9904}.
 \par
 The main idea is that by measuring the differential cross-section
%
 \begin{eqnarray}
 \frac{\der^2\sigma_{{\rm e}\gamma\rightarrow{\rm e}{\rm X}}}{\der x\der\qsq}
 &=& \frac{2\pi\aemsq}{x\,Q^{4}}
     \left[\left( 1+(1-y)^2\right) \ftxq - y^{2} \flxq\right]\, ,
 \label{eqn:DIS}
 \end{eqnarray}
%
 the photon structure function \ft\ is obtained.
 Here \qsq\ is the absolute value of the four momentum squared 
 of the virtual photon, $\gamma^{\star}$; $x$ and $y$ are the usual 
 dimensionless variables of deep inelastic scattering 
 and \aem\ is the fine structure constant.
 \par
 In the region of small $y$ studied ($y\ll 1$) the contribution 
 of the term proportional to the longitudinal structure function
 \fl\ is small and it is usually neglected.
 In leading order \ft\ is proportional to the parton content,
 $f_{i,\gamma}$, of the photon,
 $\ft=x\sum\,{\rm e}_{\rm q}^2\,(f_{\rm q,\gamma}+f_{\bar{\rm q},\gamma})$,
 where the sum runs over quarks q and antiquarks $\bar{\rm q}$ of
 charge ${\rm e}_{\rm q}$.
 The hadronic structure function \ft\ receives contributions both from the
 point-like and the hadron-like part of the photon structure.
 The point-like part can be calculated in perturbative QCD, whereas the
 hadron-like part is usually described based on the Vector Meson Dominance 
 model.
 \par
 Because the energy of the quasi-real photon is not known, the value of 
 $x= \qsq/ (\qsq+\psq+\wsq)$ has to be derived by measuring the
 invariant mass $W$ of the hadronic final state, which is a source
 of significant uncertainties, especially at low values of $x$, 
 and makes measurements of \ft\ mainly systematics limited.
 \par
 If both photons are virtual, Eq.~(\ref{eqn:DIS}) gets more complicated,
 however, in the region $\qsq\gg\psq\gg\Lambda^2$,
 where $\Lambda$ is the QCD scale, an effective structure 
 function \feff\ of virtual photons can be determined~\cite{NIS-9904}.
 \par
 This review concentrates on the recent developments to reduce the 
 systematic uncertainties, the newly evaluated radiative corrections
 to \ft, and the latest measurements of \ft\ and \feff.
%
%
\section{New developments in the analysis}
\label{sec:news}
 In previous analyses it had been seen that the dominant error in the 
 measurement of \ft\ at low $x$ is the imperfect modelling of the 
 hadronic final state by the Monte Carlo programs.
 To reduce this error two approaches have been taken. 
 Firstly, the LEP experiments have measured distributions of the hadronic
 final state corrected for detector effects~\cite{FIN-9901}.
 For large regions in most of the distributions studied
 the results of the different experiments agree with one another,
 and consequently the results have been combined while using the spread
 of the measurements as an estimate of the systematic uncertainty.
 Significant differences are found~\cite{FIN-9901}
 between the combined data and 
 the predictions of the HERWIG and PHOJET Monte Carlo models. 
 Therefore the combined LEP data serve as an important input to improve
 on the Monte Carlo models.
 \par
 Secondly, several experiments have used improved unfolding methods
 to reduce the sensitivity of the result to the different predictions.
 The main idea is the following.
 If one assumes that the structure function \ft\ is independent of the 
 fragmentation of the hadronic final state, then, in the one-dimensional
 unfolding, using the variable $x$, the result is independent of the actual
 shape of the input distribution function $\fpar(x)$ used in the unfolding,
 and only depends on the transformation $A(\xvis,x)$ between the  
 value of $x$ and the measured value \xvis.
 This transformation partly depends on the Monte Carlo model
 used, but also to a large extend on the detector capabilities which
 are independent of the chosen model.
 By using a second variable, $v$, the same argument applies
 to this variable.
 Now the result is largely independent of the joint input distribution
 function $\fpar(x,v)$ and only the transformation
 $A(\xvis,\vvis,x,v)$ matters.
 Because only the transformation of $v$ but not its predicted distribution
 affects the unfolding result, part of the dependence on the Monte Carlo
 model is removed.
 \par
%
\begin{figure}[thb]\unitlength 1pt
\begin{center}
\begin{picture}(0,0)(0,0)
 \put(026,175){ALEPH $\,\,\,\qzm=56.5\,\gevsq$}
 \put(200,175){L3}
\end{picture}
{\includegraphics[width=0.49\linewidth]{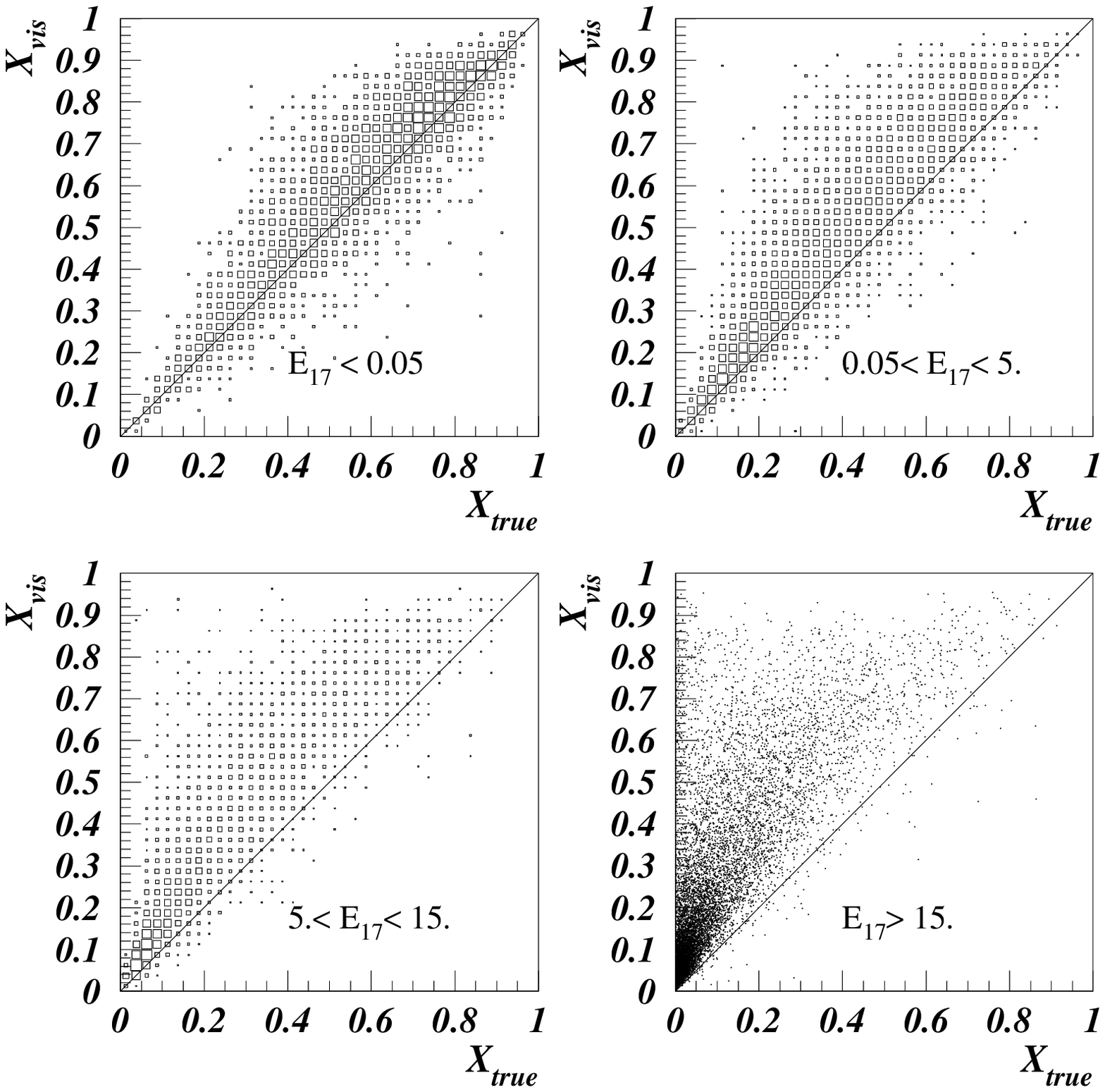}}
{\includegraphics[width=0.49\linewidth]{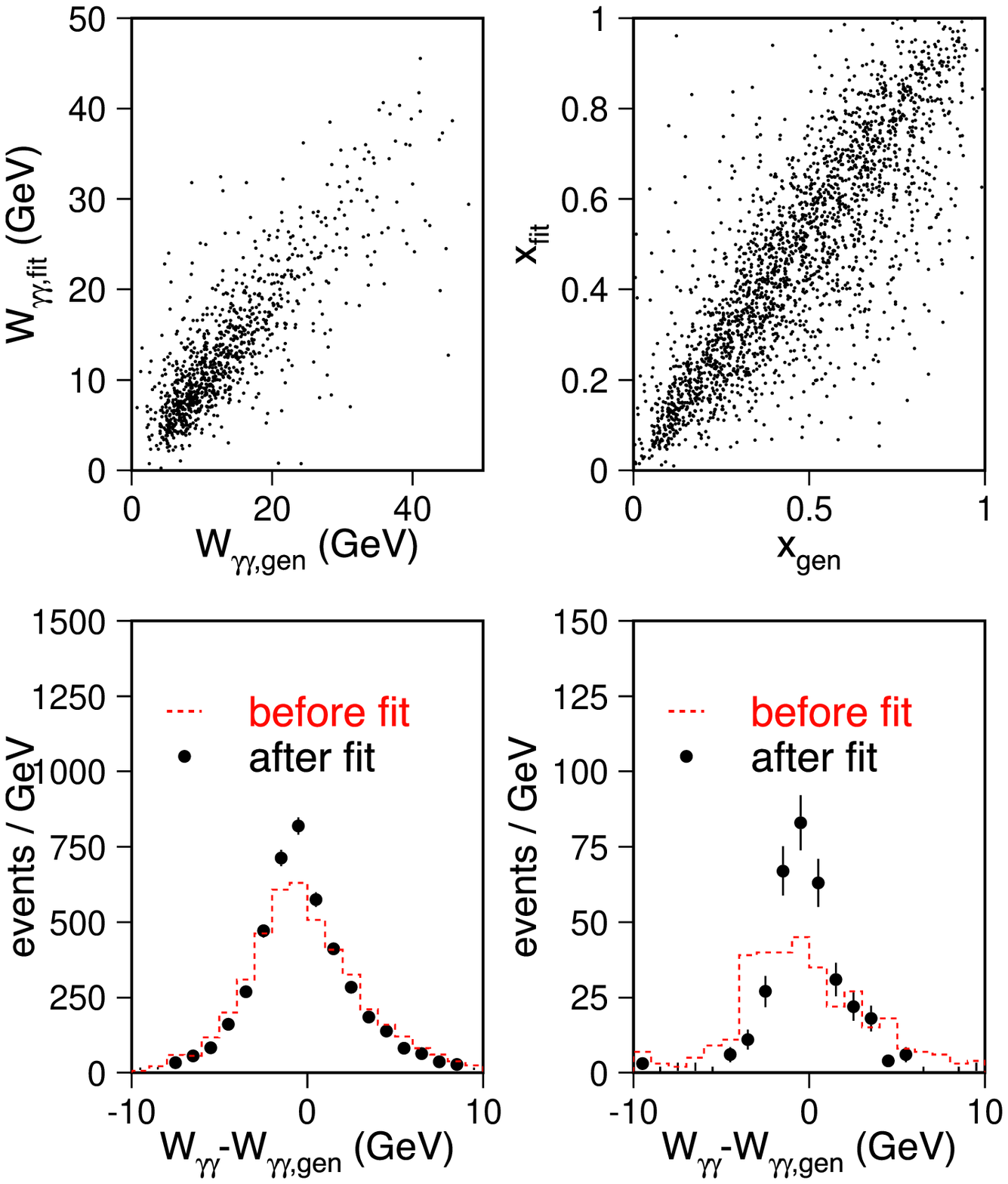}}
\caption{\label{fig01}
         Correlation between generated and measured quantities from ALEPH
         and L3.
        }
\end{center}
\end{figure}
%
 A quantity which has proven to be very useful as the second unfolding
 variable is the forward energy.
 As an example from ALEPH~\cite{ALEPHNEW},
 Figure~\ref{fig01}(left) shows the correlation between 
 $x$ and \xvis\ at $\qzm=56.5$~\gevsq\ for several
 bins of the energy observed in the forward region below a polar angle 
 of 17 degrees, denoted with $E_{17}$. 
 If very little energy is observed under small polar angles, which means
 the hadronic system is well contained in the central detector, there is
 a good correlation between $x$ and \xvis.
 In contrast, for large values of the forward energy the correlation 
 severely deteriorates.
 Using two-dimensional unfolding~\cite{2D} the result for \ft\ is
 almost independent of the largely different predicted distributions 
 of the forward energy.
 \par
 At large \qsq\ more transverse momentum is transferred to the hadronic
 system which therefore is better contained in the detector.
 It has been shown by L3 that in this case already a kinematic fit using
 energy momentum conservation gives a good correlation between the
 generated and measured hadronic invariant mass~\cite{L3C-0001}.
 Based on a generator simulating the quark parton model (QPM) prediction,
 this is demonstrated
 in Figure~\ref{fig01} for $\qzm=120$~\gevsq, for quasi-real target
 photons and also for virtual target photons with $\pzm=3.7$~\gevsq
 (lower right).
 Already without the kinematic fit an acceptable resolution is found
 for quasi-real target photons (lower left), which improves after
 the fit procedure has been applied leading to a good correlation (upper). 
 For the virtual target photons the initial resolution is somewhat worse
 and consequently the fit yields a bigger improvement (lower right). 
 \par
%
\begin{figure}[thb]\unitlength 1pt
\begin{center}
{\includegraphics[width=0.8\linewidth]{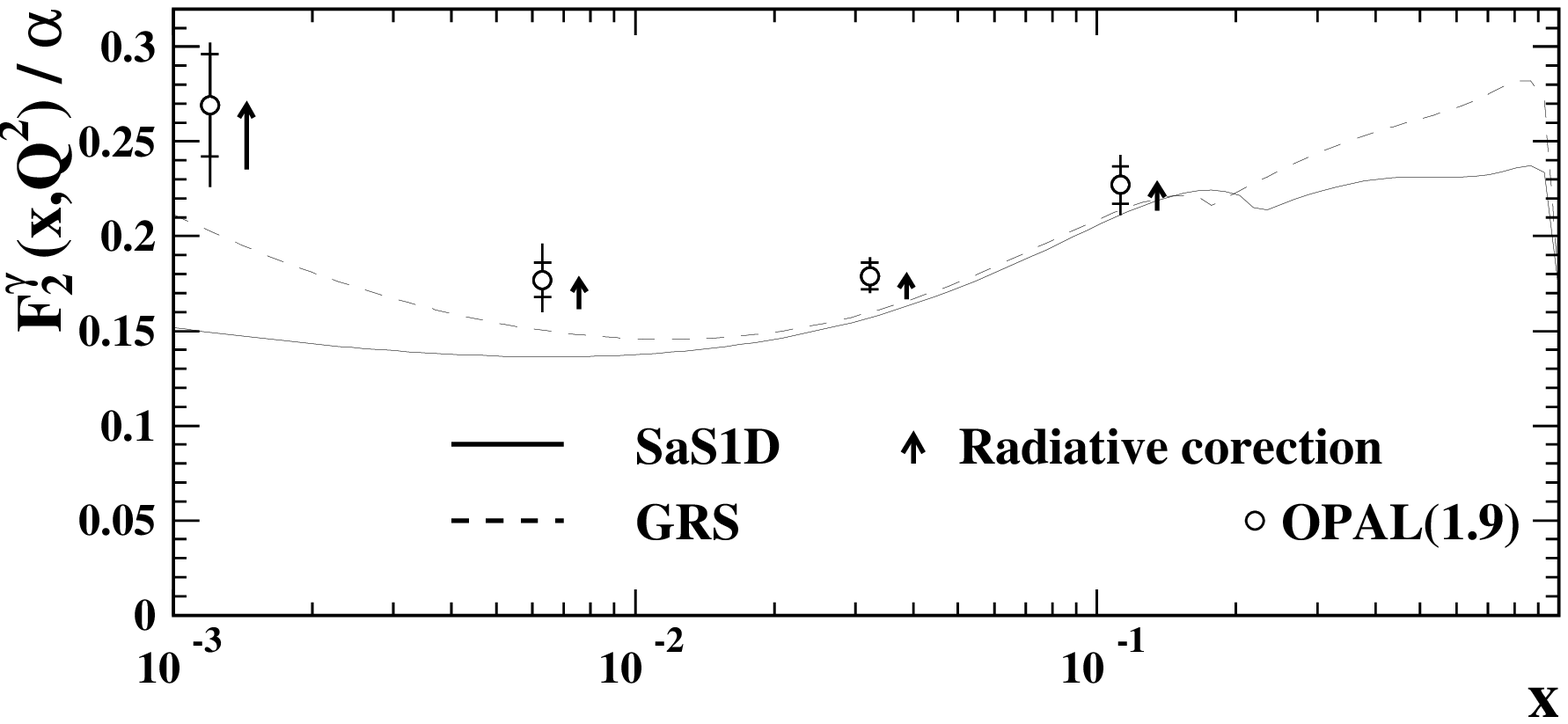}}
\caption{\label{fig02}
         The effect of radiative corrections on \ft\ for OPAL
         at $\qzm=1.9$~\gevsq.
        }
\end{center}
\end{figure}
%
 Usually QED radiative corrections to Eq.~(\ref{eqn:DIS}) are
 neglected in the analysis of \ft. 
 For this conference, OPAL for the first time presented a 
 measurement corrected for this effect based on the prediction of the
 RADEG~\cite{RADEG} program, which takes into account
 initial state radiation from the deep-inelastically scattered
 electron and the Compton scattering process. 
 It has been found that the radiative corrections are $x$-dependent 
 and largest at small values of $x$, such that the shape of \ft\ is
 changed when these corrections are applied~\cite{OPALNEW}.
 An example of the size of the corrections is shown in Figure~\ref{fig02},
 where the arrows connect the values of \ft\ before and after 
 the correction.
 With the present level of accuracy of the measurements the corrections
 are comparable to the statistical precision of the OPAL data.
 \par
 A second effect, which is usually not corrected for is
 the predicted suppression of \ft\ due to the 
 fact that the quasi-real target photon is slightly off-shell.
 This \psq\ suppression is theoretically uncertain and the predictions 
 vary by as much as a factor of two at low values of $x$.
 Therefore, at present, this correction should not be applied to the data, 
 in order not to bias the experimental result towards a particular 
 theoretical model~\cite{NIS-9904}.
%
%
\section{Recent Measurements}
\label{sec:last}
%
\begin{figure}[thb]\unitlength 1pt
\begin{center}
{\includegraphics[width=0.8\linewidth]{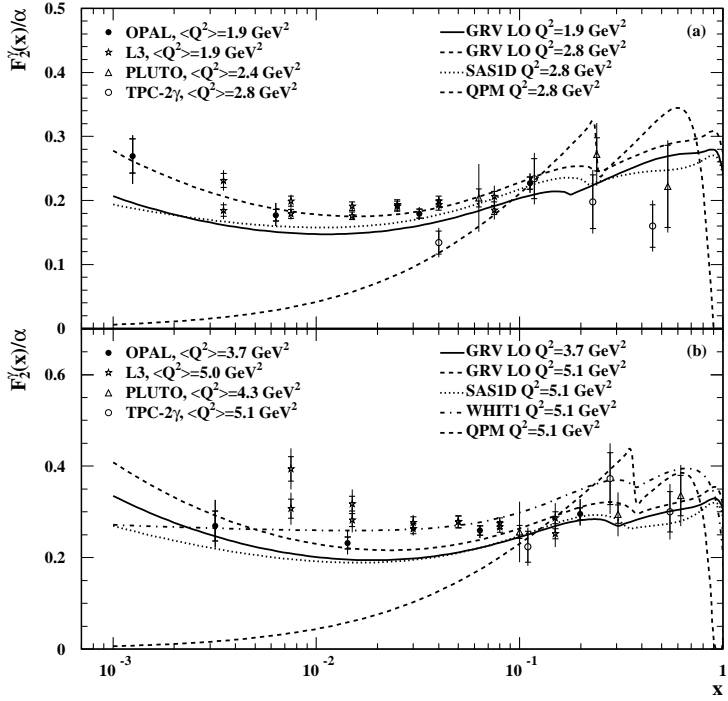}}
\caption{\label{fig03}
         New measurements of \ft\ from OPAL.
        }
\end{center}
\end{figure}
%
%
\begin{figure}[thb]\unitlength 1pt
\begin{center}
{\includegraphics[width=0.8\linewidth]{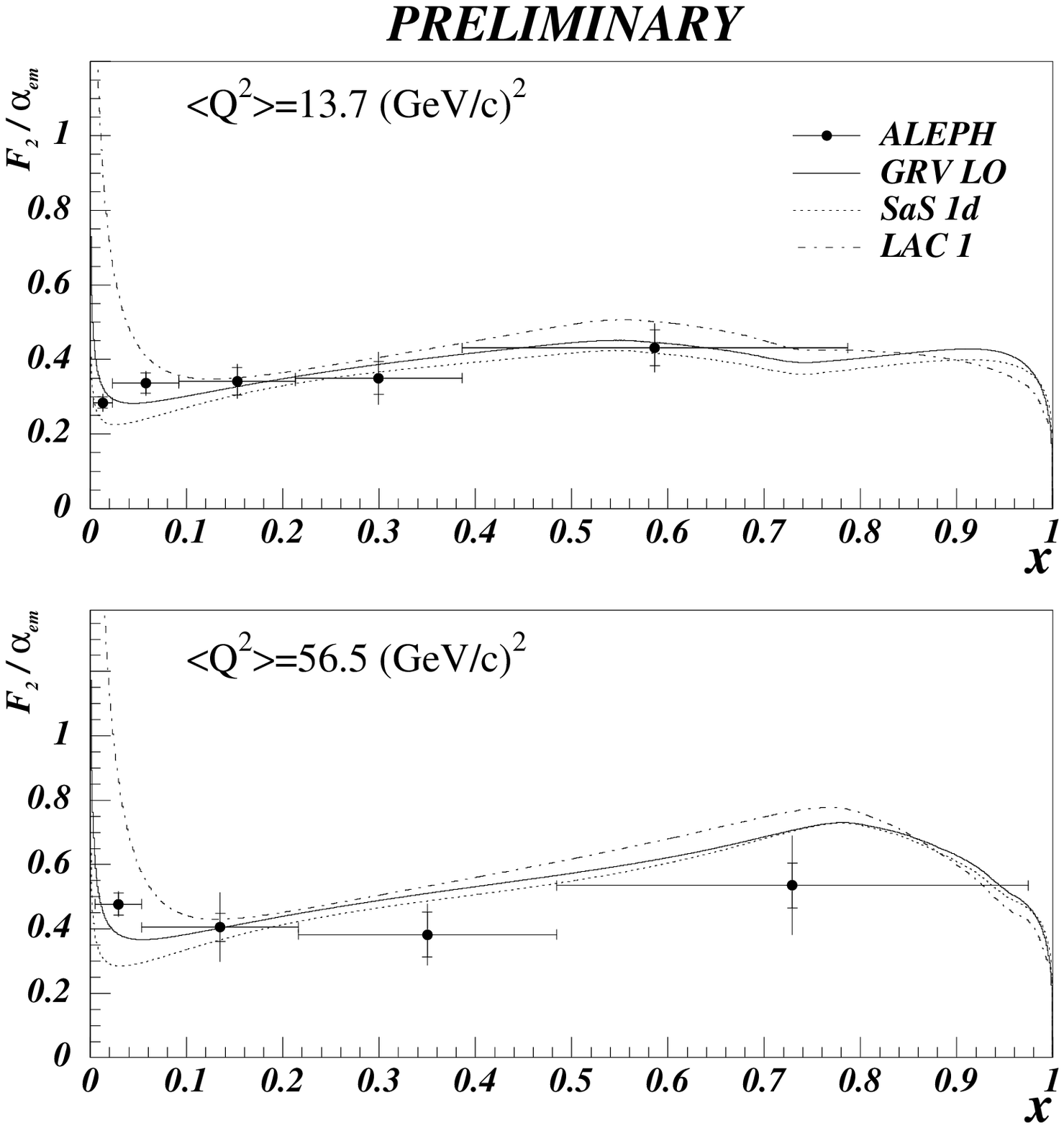}}
\caption{\label{fig04}
         New measurements of \ft\ from ALEPH.
        }
\end{center}
\end{figure}
%
 Several new or recently finalised measurements have been presented at 
 this conference.
 OPAL has updated and extended the measurements concentrating on the
 low $x$ behaviour of \ft\ for \qzm\ values ranging from 1.9 to 
 17.8~\gevsq, examples of which are shown in Figure~\ref{fig03}. 
 By using improved Monte Carlo models, improved reconstruction
 techniques to measure the visible invariant hadronic mass, and
 two-dimensional unfolding, the measurement errors have been considerably 
 reduced~\cite{OPALNEW}, compared to the published analysis.
 \par
 In general the shape of the GRV LO parameterisation is consistent 
 with the OPAL data in all the accessible $x$ and \qsq\ regions.
 The normalisation is also consistent with the data, except
 at the lowest scale, $\qzm=1.9$~\gevsq\ Figure~\ref{fig03}(a),
 where GRV is too low.
 Within the precision of the OPAL measurement, the description of the data 
 by SaS1D LO is of similar quality as for GRV LO.
 Also for L3 the shape of the GRV LO parameterisation is consistent,
 however, in this case GRV lies below the data at 
 $\qzm=5$~\gevsq\ Figure~\ref{fig03}(b).
 The LEP results extend the reach at low $x$ compared to measurements
 of \ft\ performed at lower \epem\ centre-of-mass energies.
 The results from PLUTO nicely agree with the LEP data at high values of
 $x$, whereas the shape of TPC/$2\gamma$ is largely different from 
 all other measurements.
 \par
 The new results on \ft\ from ALEPH are shown in Figure~\ref{fig04}. 
 Based on 52.9~\invpb of data, taken at \epem\ centre-of-mass energy of 
 183~\gev, \ft\ has been obtained in two \qsq\ ranges of
 $7 \leq \qsq \leq 24$~\gevsq\ and $17 \leq \qsq \leq 200$~\gevsq\ 
 with average values as indicated in the figure.
 The inner error bars represent the statistical errors 
 while the outer ones include systematical uncertainties,
 mainly coming from the remaining model dependence and the details of the 
 smoothing and regularisation technique used in the unfolding procedure. 
 The new ALEPH result clearly disfavours the strongly rising \ft\ 
 prediction from LAC1.
 The very same effect, namely that predictions of steeply rising \ft\
 at low $x$, driven by large gluon distribution functions of the photon
 are disfavoured by the data, had been seen previously~\cite{NIS-9904}.
 \par
%
\begin{figure}[thb]\unitlength 1pt
\begin{center}
{\includegraphics[width=0.49\linewidth,clip]{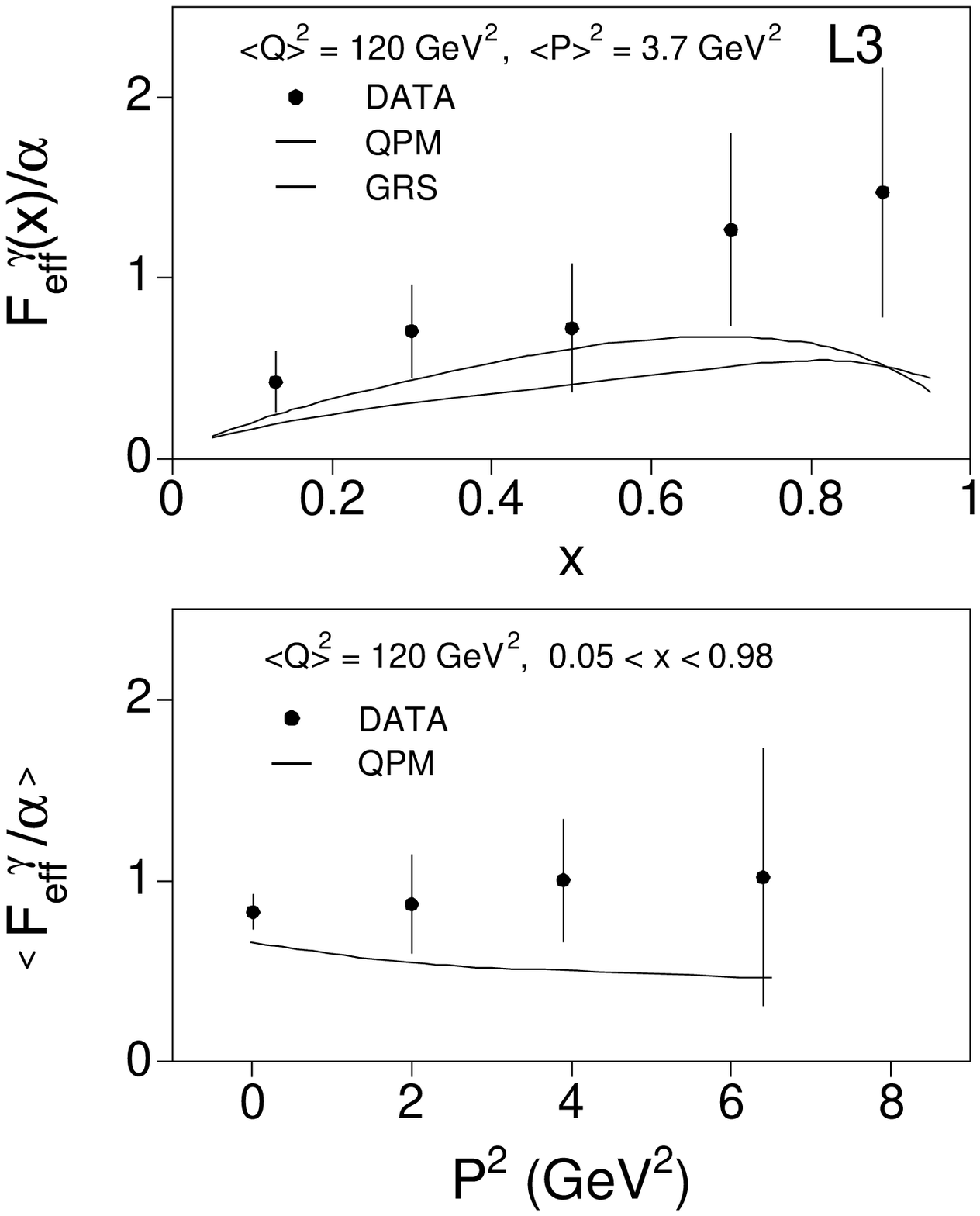}}
{\includegraphics[width=0.49\linewidth,clip]{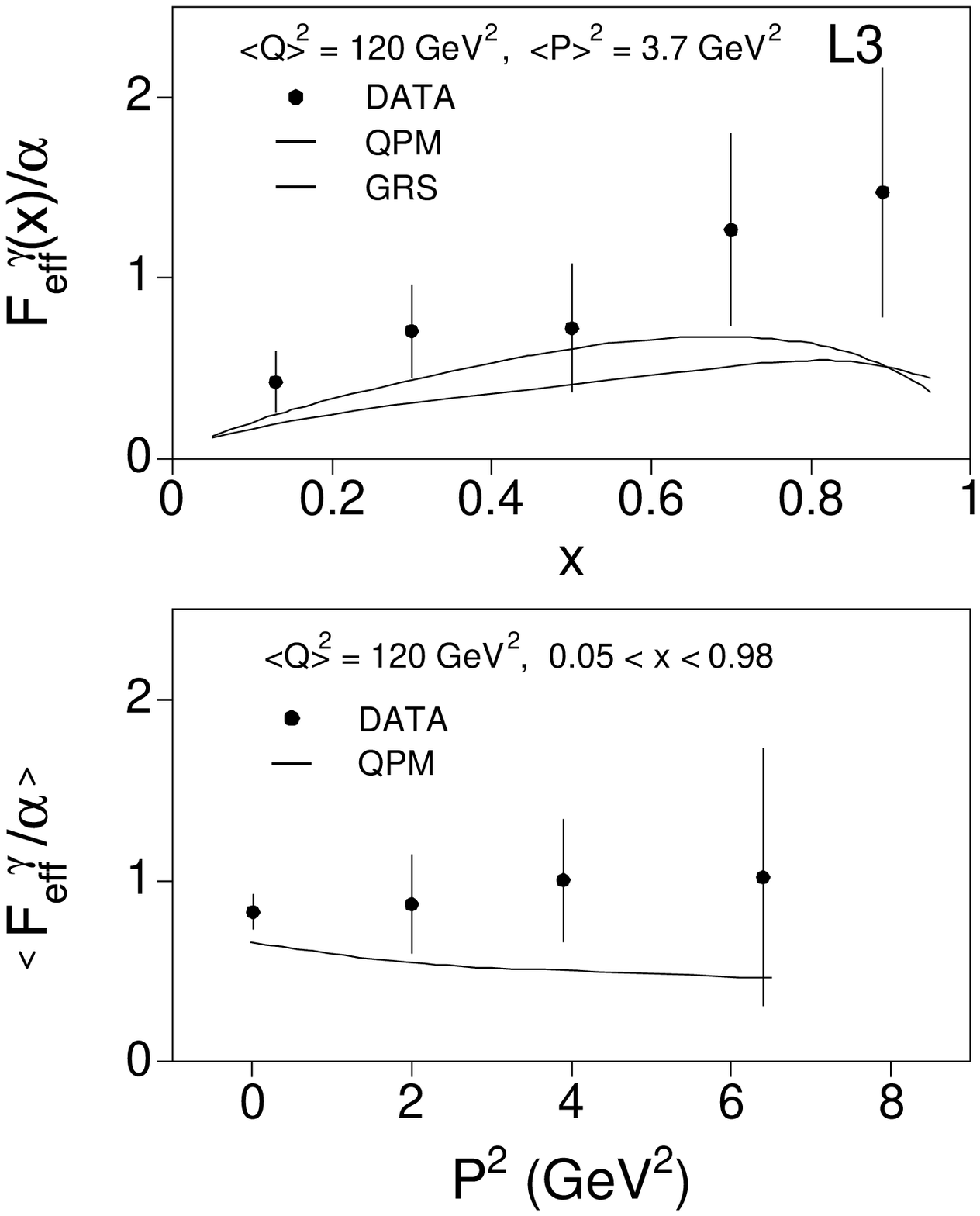}}
\caption{\label{fig05}
          Measurement of \feff\ from L3.
        }
\end{center}
\end{figure}
%
 L3 has finalised the results for $\qzm = 120$~\gevsq\
 based on LEP data taken at \epem\ centre-of-mass energies around the
 mass of the $Z$ boson~\cite{L3C-0001}.
 The structure function \ft\ is not described by the quark parton model
 for $x<0.4$, where the hadron-like component is expected to be largest.
 In this region the data are even higher than the predictions
 of several parametrisations of \ft\ which contain a hadron-like
 contribution~\cite{L3C-0001}.
 \par
 In addition, the structure function \feff\
 was measured for $\qzm = 120$~\gevsq\ and $\pzm = 3.7$~\gevsq, 
 thereby ensuring $\qsq\gg\psq\gg\Lambda^2$, Figure~\ref{fig05}.
 As in the case of the PLUTO result~\cite{PLU-8405}, the QPM 
 prediction is too low compared to the data.
 Also the GRS prediction falls short with respect to the data. 
 However, this may be expected, because the GRS prediction only 
 contains the contribution from transverse virtual target photons.
 The QPM prediction of the \psq\ evolution of \feff\ is consistent in
 shape with the data, but too low, with the most significant difference
 stemming from \ft\ at $\psq=0$ and for $x<0.4$~\cite{L3C-0001}.
 The measurement at $\psq>0$ cannot rule out the quark parton model
 prediction, although the data are consistently higher.
 For more detailed comparisons to be made the full statistics of the LEP2
 programme has to be explored.
%
%
\section{The present status of the measurements}
\label{sec:status}
 In the present investigations of the photon structure function \ft\
 two distinct features of the photon structure are studied.
 Firstly, the shape of \ft\ is measured as a function of $x$ at fixed \qsq.
 Particular emphasis is put on measuring the low $x$ behaviour of
 \ft\ in comparison to the proton structure function obtained at HERA.
 Predictions of strongly rising \ft\ at low $x$ are disfavoured by the
 data which show a rather flat behaviour of \ft\ at low $x$.
 \par
%
\begin{figure}[thb]\unitlength 1pt
\begin{center}
{\includegraphics[width=0.95\linewidth]{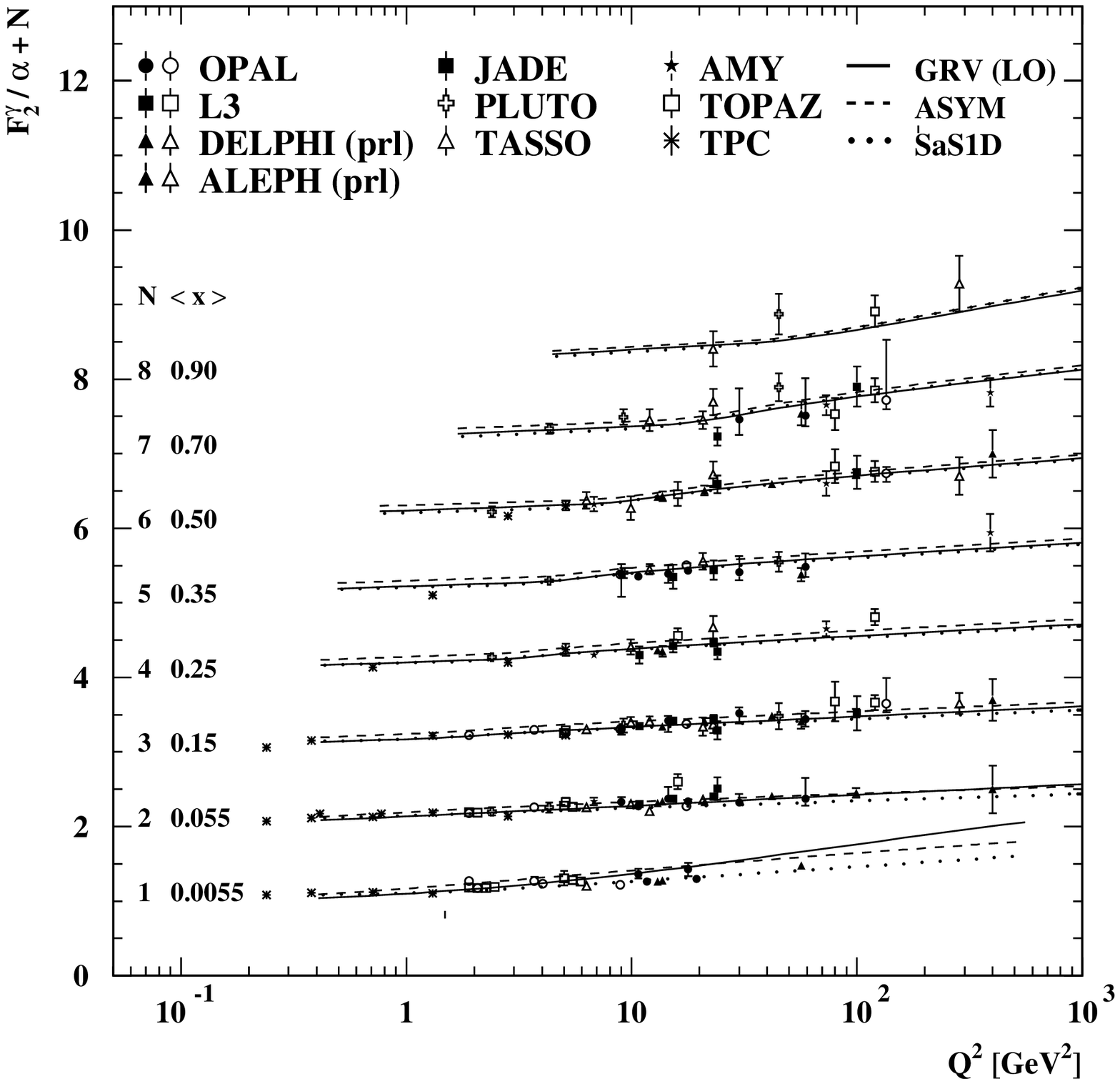}}
\caption{\label{fig06}
         The \qsq\ evolution of \ft\ in bins of $x$.
        }
\end{center}
\end{figure}
%
 Secondly, the evolution of \ft\ with \qsq\ is investigated.
 The present status of the measurements, including those discussed above,
 is shown in Figure~\ref{fig06}.
 The positive scaling violation predicted by QCD for all values of $x$
 is clearly seen. 
 In addition, it has been seen~\cite{NIS-9904} that the slope of \ft\ 
 increases for increasing values of $x$, and that the measurements can 
 be described by an augmented asymptotic prediction~\cite{WIT-7701}
 (ASYM) for \ft\ as described in more detail in~\cite{NIS-9904}.
%
%
\section{Conclusions}
\label{sec:concl}
 The measurement of the hadronic structure of the photon is an active 
 field of research.
 The newly available results indicate that the systematic error of the
 measurements can be largely decreased when two-dimensional unfolding
 and improved Monte Carlo models are used.
 With the analysis of the full data taken within the LEP2 programme, 
 considerable improvement of our understanding of the hadronic 
 structure of the photon can be expected.
%
%
\section*{Acknowledgments}
 We wish to thank the organisers of this interesting workshop
 for the fruitful atmosphere they created throughout the meeting. 
 One of the authors (C.~G.) wants to thank Klaus Affholderbach for his 
 help in preparing the talk.
%
%

\end{document}